\documentclass[aps,prl,showpacs,preprintnumbers,nofootinbib,preprint]{revtex4}

\usepackage{bbm}
\usepackage{latexsym}
\usepackage{dcolumn}
\usepackage{amsfonts,amssymb}
\usepackage{graphicx,epsfig}
\usepackage{psfrag}
\usepackage{amsmath,amssymb}

\usepackage[T1]{fontenc}

\def\vp{\varphi}

\def\beq{\begin{equation}}
\def\eeq{\end{equation}}
\def\br{\begin{eqnarray}}
\def\er{\end{eqnarray}}
\def\benu{\begin{enumerate}}
\def\eenu{\end{enumerate}}
\def\nn{\nonumber} 

\def\l{\left}
\def\r{\right}    

\def\Hbar{\mathcal H}

\def\MPl{M_{_{\rm Pl}}^2}

\newsavebox{\blambox}\savebox{\blambox}[0.6em]{$\lambda\!\!\!$\raisebox{0.5em}
{$\neg$}}\newcommand{\blambda}{\usebox{\blambox}}
\newsavebox{\bFox}\savebox{\bFox}[0.6em]{$F\!\!\!\!$\raisebox{0.5em}
{$\neg$}}\newcommand{\bF}{\usebox{\bFox}}
\newsavebox{\bxibox}\savebox{\bxibox}[0.6em]{$\xi\!\!\!$\raisebox{.5em}
{$\neg$}}

\newcommand{\bphi}{\overline{\varphi}}
\newcommand{\dphi}{\delta \varphi}
\newcommand{\Hm}{\mathcal{H}}

\newcommand{\Fpt}{{\mathcal{F}}}
\newcommand{\Fpteps}{{\mathcal{F}}_{\varepsilon}}

\begin{document}
\title{Modified scalar and tensor spectra in spinor driven inflation}
\author{Damien Gredat$^{1,2}$, S.~Shankaranarayanan$^2$\footnote{Current address: School of Physics, Indian Institute of Science Education and Research (IISER), CET Campus, Trivandrum 695016, India}}
\affiliation{$^1$ Physique Fondamentale et appliqu\'ee, 
Universit\'e Paris-Sud 11, 91405 Orsay cedex, France \\
$^2$ Institute of Cosmology and Gravitation, University of Portsmouth,
Portsmouth P01 2EG, U.K.} 
\email{damien.gredat@u-psud.fr,shanki@iisertvm.ac.in}

\begin{abstract}

One of the firm predictions of single-scalar field inflationary
cosmology is the consistency relation between scalar and tensor
perturbations.  It has been argued that such a relation, if
observationally verified, would offer strong support for the idea of
inflation. In this letter, we critically analyze the validity of the
consistency relation in the context of spinor driven inflation. The spinflaton --
a condensate of the Elko field --- has a single scalar degree of
freedom and leads to the same acceleration equation as the
inflaton. We obtain the perturbation equations for the Einstein-Elko
system and show that the scalar perturbations are purely adiabatic and
the sound speed of the perturbations is identically one. We obtain the
generalized Mukhanov-Sasaki equation for the spinor driven inflation and show
that, in the slow-roll limit, the scalar and tensor spectra are nearly
scale-invariant. We also show that spinor driven inflation naturally predicts
running of spectral indices and the consistency relations for the
spectra are modified.
\end{abstract}
\maketitle

Observations of the cosmic microwave background (CMB) temperature
variations, combined with measurements of the large scale structure
(LSS), strongly favor a period of accelerated expansion in the early
universe --- the inflationary paradigm \cite{Komatsu2008}. The
inflationary epoch magnifies quantum fluctuations into primordial
classical perturbations that leave an imprint as anisotropies in the
CMB and seed the LSS. Slow-roll inflation predicts that the primordial
perturbations are scale-invariant.

Although the current observations lend great support to the
inflationary paradigm, inflation has several theoretical problems to
address, including, the hierarchy problem, reheating problem and
trans-Planckian problem
\cite{Brandenberger2008}. The questions often seem related and linked
to the fundamental issue: What is the nature of the inflaton (field 
that dominates during inflation) \cite{Brout2001}?

Specifically, it is still unclear whether the inflaton $(\vp_{\rm c})$
which appears in the acceleration equation:
{\small
\beq
\label{eq:acceleration}
\frac{\ddot{a}(t)}{a(t)} = \frac{1}{3 \MPl} \l[V(\vp_{\rm c}) - 
\dot{\vp_{\rm c}}^2\r] 
\qquad \MPl \equiv (8 \pi G)^{-1}
\eeq
}
\hspace*{-7pt}(where $c = \hbar = 1$, $V(\vp_{\rm c})$ is the inflaton 
potential) is a fundamental scalar field or an effective description
of several fields. Although, originally, the inflaton was considered
to be a fundamental scalar field, there is more theoretical evidence
that it can at best be an effective one. The different approaches that
have been adopted in the literature can be classified into three
categories:
(i) f(R) gravity models \cite{Starobinsky1980}, 
(ii) Multiple scalar fields \cite{GrootNibbelink-Vant:2001}, 
(iii) Higher spin fields, for example, vector fields
\cite{Ford1989}.

However, the possibility that the scalar field is a condensate of
spinors has received less attention in the literature. This can be
attributed to one or several of the following reasons:
(i) Difficult to obtain a consistent solution of the Einstein-Dirac
system for an FRW background \cite{Saha2001b,Boyle2002}.
(ii) Standard model spinors do not lead to $60$ e-foldings of inflation
\cite{Armendariz-Picon2003,Boehmer2007i,Boehmer2008}.
(iii) Deriving the metric perturbations and the perturbations of the
spinorial part of the field is difficult. In fact, to our knowledge,
such an analysis has not been done in the literature.

In this work, we show that the scalar field $(\vp)$ as a fermion
condensate is a viable model of inflation. The physical picture is the
following: Like in superconductivity, the spinors form a condensate
which dominates in the early universe. In the FRW background, the
symmetry demands that the spinors depend on a single scalar degree of
freedom (which we refer to as the {\it spinflaton}). Interestingly,
the spinflaton leads to the same acceleration equation
(\ref{eq:acceleration}), where $V(\vp)$ is the potential of the
condensate. For specific form of metric perturbations, the
perturbations effectively has a single degree of freedom.

The fermions we consider, in this work, are non-standard Wigner-class
spinors. These were recently discovered by Ahluwalia-Khalilova and
Grumiller \cite{Ahluwalia2005}, and referred to these spinors as Elko
[$\lambda$]. Elko differs from the standard model spinors
\cite{Ahluwalia2005}:
(i) Elko is an eigen-spinor of charge conjugation operator and has
mass dimension one.
(ii) They obey the unusual property $({\rm CPT})^2 = -\mathbbm{1}$. 
(iii) The dominant interactions of Elko with standard matter field
take place through the Higgs doublet or via gravity. These spinors 
were originally introduced in Refs. \cite{Ahluwalia2005} as a plausible 
dark matter candidates.

In this work, we show that:
(i) As in single scalar field inflation, the tensor and scalar 
perturbations are generated from the spinflaton  
are not independent, leading to modified consistency relations. 
(ii) The perturbations generated during the inflation are purely
adiabatic on large scales.
(iii) The scalar and tensor spectra are nearly scale invariant and
lead to running of indices.

\noindent {\it Set-up:} We consider the following 4-dimensional action:
{\small
\beq
\label{eq:action}
S = \int d^4x \sqrt{-g} \l[\frac{\MPl}{2} \, R + \mathfrak{L}_{\rm elko} \r]
\eeq
}
\hspace*{-7pt} 
where the Lagrangian for the Elko field is given by \cite{Ahluwalia2005}
{\small
\beq
\label{eq:ElkoLag}
\mathfrak{L}_{elko}= 
\frac{1}{2} 
\l[\frac{1}{2} g^{\mu \nu}(\mathfrak{D}_\mu \blambda \mathfrak{D}_\nu \lambda 
+ \mathfrak{D}_\nu \blambda \mathfrak{D}_\mu \lambda) \r] 
- V(\blambda \lambda) \, .
\eeq
}
\hspace*{-6.5pt}$V(\blambda \lambda)$ is the potential of the Elko field, $\blambda$
is the elko dual [defined in Eq. (\ref{eq:elkosform})] and
the covariant derivatives are 
{\small
\beq
\mathfrak{D}_\mu \lambda = 
(\overrightarrow{\partial}_\mu + \Omega_\mu)\,\lambda(x)~;~
\mathfrak{D}_\mu\blambda = 
\blambda(x) \, (\overleftarrow{\partial}_\mu - \Omega_\mu) \, ,
\eeq
}
\hspace*{-7pt}where the tangent space connection\footnote{The vierbeins
$(e^a_{\mu})$ satisfy the orthonormality relations $g_{\mu
\nu}=e^a_\mu e^b_\nu \eta_{ab}$, $\eta^{ab}=e_\mu^a e_\nu^b g^{\mu
\nu}$. In this work, $\Gamma^\mu(\gamma^a)$ corresponds to spinors in
curved(Minkowski) space-time, Greek (Latin) alphabets correspond to
the space-time (internal) indices, spinors in curved and Minkowski
space-time are related by $\Gamma^\mu=e^\mu_a \gamma^a$ and satisfy
the extended Clifford algebra $\{\Gamma^{\mu}, \Gamma^{\nu}\} = 2
g^{\mu\nu} \mathbbm{1}$, $\{\gamma^{a}, \gamma^{b}\} = 2 \eta^{ab} \mathbbm{1}$.} $\Omega_\mu$ is  
{\small
\beq
\Omega_\mu= - \frac{i}{2}\, \l( e_{\alpha a}e^\nu_b\Gamma^\alpha_{\mu \nu}
- e^\nu_b\partial_\mu e_{a \nu} \r) S^{ab}, \quad 
S^{ab} = \frac{i}{4} [\gamma^a,\gamma^b]
\eeq
}
\hspace*{-7pt} is the generator of the Lorentz transformations
and $\gamma$'s are the Dirac matrices in the Weyl representation. 

As mentioned, Elko is an eigen-spinor of the charge conjugation
operator and has the following form
\cite{Ahluwalia2005,Ahluwalia:2008xi}:
{\small
\begin{equation}
\label{eq:elkosform}
\lambda(x) =\left(\begin{array}{c} 
              \pm \sigma_2 {\phi^{(1)}_{\rm L}}^* \\ 
              \phi^{(1)}_{\rm L} 
              \end{array} \right)
\qquad 
\blambda(x) = i \left({\phi_{\rm L}^{(2)}}^{\dagger} 
                   \pm {\phi^{(2)}_{\rm L}}^{\dagger} \sigma_2 
             \right)
\end{equation}
}
\hspace*{-7pt} where $\sigma_2$ is Pauli matrix,
$+(-)$ correspond to the (anti-)self conjugate spinor with respect to
the charge conjugation operator, and subscript $L$ refers to
left-handed spinor. [Similar definitions hold for right-handed
spinor.]

The following points are worth noting regarding the form of Elko:
(i) $\phi_{\rm L}^{(1)}, \phi_{\rm L}^{(2)}$ refer to the two
different choices of helicities. Unlike standard spinors, the two
helicities are to be treated differently, which is due to the fact
that Elko is not the eigen spinor of the helicity operator.
(ii) The form of Elko (\ref{eq:elkosform}) is different from the
standard model spinors. In order to have a real norm, Elko duals
$(\blambda)$ are constructed differently compared to the Dirac adjoint, i.e
{\small
\beq
\label{eq:Enorm}
\blambda \lambda = \varphi^2(x) \, , \qquad  
\varphi~\mbox{is a real scalar} \, .
\eeq
}
\hspace*{-7pt}(iii) Elkos have four complex functions; not all of them 
are independent and related by the above constraint.

Our aim is to study the evolution of perturbations of the Elko during
spinor driven inflation. To linear order in fluctuations (and neglecting vector
modes), the line element --- in the longitudinal gauge --- for a
spatially flat FRW background can be written as
\cite{Kodama-Sasa:1984}:
{\small
\beq
\label{eq:LinearFRW} 
{\rm d}s^2 = 
a^2 \, \{(1 + 2 \Phi ){\rm d}\eta ^2  
- [(1- 2 \Psi )\delta _{ij} + h_{ij}]{\rm d}x^i{\rm d}x^j\}\, , 
\eeq
}
\hspace*{-6.5pt} where $\Phi, \Psi$ are the Bardeen potentials and 
represent the scalar sector, and the traceless and transverse tensor
$h_{ij}$ ($h_i{}^i=h_{ij}{}^{,j}=0$), represents the tensor sector,
i.e. the gravitational waves. $\eta = \int [dt/a(t)]$ is the conformal
time, the Hubble rate is $H \equiv \dot{a}/a = {\cal H} a$, ${\cal H}
\equiv {a}'/a$ and a prime refers to derivative with respect to
$\eta$. In this work, we present the key results and their implications, 
the detailed calculations will be presented in Ref. \cite{DamienShanki:2008b}.
\par
\par
 
\noindent {\it Homogeneous and isotropic background:} First, let us study the 
evolution of Elko in the FRW background. [Variables with an over-bar
are evaluated in the FRW background.] Using the following form for the
background Elko
{\small
\beq
\label{eq:Elko-bg}
\overline{\lambda} = \frac{\overline{\varphi}(\eta)}{\sqrt[4]{12}}
\left( \begin{array}{c} 
       -\alpha_{_1}  e^{i \pi/4} \\ 
        \alpha_{_2} \frac{i}{\sqrt{2}} \\ 
        \alpha_{_2} \frac{1}{\sqrt{2}} \\ 
        \alpha_{_1} e^{i \pi/4} 
        \end{array} \right) 
\quad \, \,
\overline{\blambda}^{\bf T} = \frac{\overline{\varphi}(\eta)}{\sqrt[4]{12}}
\left(\begin{array}{c} 
      -\alpha_{_1} e^{-i\frac{\pi}{4}} \\
      -\alpha_{_2} \frac{i}{\sqrt{2}} \\ 
       \alpha_{_2} \frac{1}{\sqrt{2}} \\ 
       \alpha_{_1} e^{-i \frac{\pi}{4}} 
     \end{array} \right) 
\eeq
}
\noindent \hspace*{-7pt} 
where $\alpha_{_1} = \alpha_{_2}^{-1} = \sqrt{1 + \sqrt{3}}$, 
the equations in the FRW background are given by:
{\small
\begin{subequations}
\br 
\label{eq:EFriedmann}
\mathcal{H}^2 = \frac{1}{3 \MPl} 
\l[\frac{{\bphi'}^2/2 
      + a^2(\eta) V(\overline{\varphi})}{1 + \Fpt}\r]&;&
\Fpt = \frac{\bphi^2}{8 \MPl}  \\
\label{eq:Eacceleration}
{\cal H}' = \frac{1}{3 \MPl} 
\l[a^2(\eta) V(\overline{\varphi}) - {\bphi'}^2(\eta)\r] & & \\
\label{eq:EKG}
\bphi''+ 2\mathcal{H}\bphi' 
+ \mathcal{G}(\bphi) + a^2(\eta)V_{,\overline{\varphi}}=0&;& 
 {\cal G}(\bphi)
= \frac{-3 \MPl}{\bphi'} \, \left[\Hm^2 \Fpt \right]'~~~~~
\er
\end{subequations}
}
\hspace*{-6.5pt} The following points are worth noting 
regarding the above results:
(i) The form of background Elko (\ref{eq:Elko-bg}) is different from
that used by Boehmer \cite{Boehmer2007i,Boehmer2008}. Unlike the one
used in Refs. \cite{Boehmer2007i,Boehmer2008}, these do not introduce
any preferred direction when studying the perturbation equations.
(iii) The background Elko (and its dual) depends on a single scalar
function $(\overline{\varphi})$ satisfying $ \overline{\blambda}
\overline{\lambda} = \overline{\varphi}^2(\eta)$. This can be
interpreted as an Elko-pair (similar to Dirac-pair)
forming a scalar condensate --- spinflaton.
(iii) As mentioned earlier, the acceleration equation for the
spinflaton (\ref{eq:Eacceleration}) is identical to
Eq. (\ref{eq:acceleration}). However, the Friedmann
(\ref{eq:EFriedmann}) and spinflaton (\ref{eq:EKG}) equations have
non-trivial corrections due to Elko. The Elko modification to the
canonical inflaton equations are determined by $\Fpt$.
(iv) Slow-roll parameters for the spinflaton are given by:
{\small
\begin{subequations}
\br
\label{eq:Eslowroll1}
\!\!\!\!\!\!\!\! & & \varepsilon \equiv - \frac{\dot{H}}{H^2} 
= \varepsilon_{_{\rm can}} [1+\Fpt(\bphi)] - \Fpt(\bphi)  \\
\label{eq:Eslowroll2}
\!\!\!\!\!\!\!\! & & \delta \equiv 
- \frac{\ddot{\overline{\varphi}}}{H \dot{\overline{\varphi}}} 
= \varepsilon  - \frac{(\varepsilon + \Fpt)'}{2 \Hbar (\varepsilon + \Fpt)} \\
%
\label{eq:Cslowroll}
\!\!\!\!\!\!\!\! & & \rm{where} ~~~\varepsilon_{_{\rm can}} = 
3 \frac{{\overline{\varphi}'}^2/2}{{\overline{\varphi}'}^2/2 
+ a^2 V(\overline{\varphi})} \, ,
\delta_{_{\rm can}} =  \varepsilon_{_{\rm can}} 
- \frac{\varepsilon'_{_{\rm can}}}{2 \Hbar \varepsilon_{_{\rm can}}} \qquad 
\er
\end{subequations}
}
\hspace*{-6.5pt}  Note that the de Sitter solution derived by Boehmer \cite{Boehmer2008} 
satisfies the above slow-roll conditions. 
(v) Assuming that the slow-roll parameters are small ($\varepsilon,
|\delta| \ll 1$), the number of e-foldings of inflation is given by:
{\small
\beq
N(\bphi) = \frac{2}{3 \MPl}\int_{\bphi_{\rm end}}^{\bphi} 
\frac{d\bphi}{(1 + \Fpt)} 
\frac{V}{V_{,\bphi} + a^{-2} \, {\cal G}(\bphi)}  \, .
\eeq
}
\hspace*{-7pt} Using the fact that ${\cal G}(\bphi)$ is strictly negative, 
we note that the potentials satisfying slow-roll for the canonical
scalar field will lead to a longer period of inflation and hence can
lead to more than $60$ e-foldings of inflation.
(v) It is important to note that, even if, $\Fpt$ is small, presence
of $1/\bphi'$ in Eq. (\ref{eq:EKG}) can introduce large, non-trivial
corrections to the perturbation equations. In the case of canonical 
scalar field, the de Sitter solution is possible only for 
$V(\varphi_{_{\rm cal}}) = constant$ and $\varphi_{_{\rm cal}} = constant$. 
In the case of Elko, the de Sitter solution obtained by Boehmer \cite{Boehmer2008} 
requires the condensate potential to be $3 q^2 \MPl + \frac{q^2}{4} \varphi^2$ 
where $q$ is a constant. As we show explicitly, the Elko corrections play a 
significant role in the power spectra. For instance, the running of spectral index 
is due to the corrections from the Elko

\noindent {\it Perturbations:}  The perturbed Einstein equations 
provide us with the equations of motions for the cosmological
perturbations. At the linear level, scalar and tensor perturbations
decouple and can be treated separately. As in standard inflation, the
tensor perturbations do not couple to the energy density and pressure
inhomogeneities. They are free gravitational waves and satisfy:
{\small
\beq
\label{eq:TensorPertu}
\mu_{_T}'' + \l(k^2 - \frac{a''(\eta)}{a(\eta)}\r)\mu_{_T} = 0 \, .
\eeq
}
\indent The main hurdle in obtaining perturbation equation for the scalar
perturbations is the perturbed stress-tensor for the Elko. Unlike the
scalar field, which has only one degree of freedom, Elko, in general,
has four complex functions; not all of these are independent and are
related by constraints\footnote{It is incomplete to study the
perturbation of the effective scalar-field equation and then derive
the Mukhanov-Sasaki equation. This is because, unlike Klein-Gordan
field, the effective scalar equation we have obtained is only valid
for the FRW background and not for any general curved
space-time.}. The most general perturbed Elko can lead to the scalar
and vector perturbations. Besides, it can also lead to a non-vanishing
anisotropic stress.

In this work,
(i) we assume that the anisotropic stress of the perturbed Elko (and
its dual) is identically zero. In other words, in the limit of $\Phi =
\Psi$, the stress-tensor of the mixed space-space components must
vanish.
(ii) We extract the scalar part of the perturbations from the $(0i)$
perturbation equation. In other words, $(0i)$ equation should only
contain a spatial derivative and the terms with out the spatial
derivatives must vanish.
(iii) Following (\ref{eq:Enorm}), the perturbed Elko must satisfy
$\overline{\blambda} \delta\lambda + \delta\blambda \overline{\lambda}
= 2 \bphi \delta\varphi$ where $\delta \varphi(x)$ is the perturbed scalar
condensate.
The above conditions ensure that the perturbation of Elko-Einstein system 
leads to a consistent linear order perturbation equations with a 
single scalar degree of freedom.

The perturbed Elko (and its dual) which satisfy all the above 
conditions is:
{\small
\beq
\label{eq:peransatz}
\delta \lambda(x^\mu)= i \, F[\gamma] \, \overline{\lambda}(\eta) 
\qquad \delta \blambda(x^\mu) = -i \, \overline{\blambda}(\eta) \, 
\l[\gamma^0 \, \bF[\gamma]^\dagger \, \gamma^0 \r]   
\eeq
}
where 
{\small
\begin{subequations}
\br
F[\Gamma] = Q_{_1} \l[ S^{\eta \mu} \, \delta\varphi_{_1}(x)
          + i \, \gamma^5 \, R^\mu \, \delta\varphi_{_2}(x) \r] \, 
           u_\mu \, ,& & \\ 
\bF[\Gamma] = 2 \sqrt{3} \, Q_{_2} \l[S^{\eta \mu} \, \delta\varphi_{_3}(x)
            + i \, \gamma^5 \, R^\mu \, \delta\varphi_{_4}(x) \r] \, 
            u_\mu \, , & &  \\
R_\mu = \frac{1}{2}\varepsilon_{\mu ij}S ^{i j} \, , 
\quad \Gamma^5 = i\sqrt{-\textbf{g}}\, 
\Gamma^0\Gamma^1\Gamma^2\Gamma^3=\gamma^5 \, , & & \\
u_\mu = \frac{1}{\sqrt{3}} \left(\begin{array}{c} 
						0 \\ 
						1 \\ 
						1 \\ 
						1 
				\end{array} \right), ~~ 
Q_{_{\ell}} =\left( \begin{array}{cccc}
 		1 & 0 & 0 & 0 \\ 
		0 & r_{_{\ell}} & 0 & 0 \\ 
		0 & 0 & r_{_{\ell}} & 0 \\ 
		0 & 0 & 0 & 1 \\
		\end{array}\right) ~~ \ell = 1, 2  & & 
\er
\end{subequations}
}
such that 
{\small
\begin{subequations}
\label{eq:Constraints}
\br
\label{eq:const1}
\delta\varphi_{_1} = \delta\varphi_{_2} 
= \delta \varphi_{_3} 
= \delta \varphi_{_4}  \, ; \,
r_{_1} = - r_{_2} = 1 \, ; ~~~ & & \\
\label{eq:const2}
\delta\varphi = \overline{\varphi} \delta\varphi_1 \, ; \,
\Phi = \Psi \, ; \,
2 \nabla \delta\varphi_1 - \sqrt{3} \partial_z \delta \varphi_2 
= 4 \nabla \Psi \, .~~~& & 
\er
\end{subequations}
}
\indent The above choice of perturbed Elko (\ref{eq:peransatz}) is inspired by
the Hedgehog ansatz \cite{Chodos1975} and has the following features:
(i) In the above expression, the partial derivative of the perturbed
field $\delta \varphi_2$ is w.r.t the coordinate axis ($z$). This
implies that the effective perturbation equations has a directional
dependence as in the case of vector inflation \cite{Ford1989}.
(ii) The perturbed Elko and its dual are related to six unknown scalar
functions $(\delta\varphi_{_1},\delta\varphi_{_2},
\delta\varphi_{_3}, \delta\varphi_{_4}, r_{_l})$ by Lorentz
boost $S^{\eta \mu}$ and the rotations $\gamma^5 R^\mu$. 
(iii)
$Q_{_{\ell}}$ scales the perturbed Elko and its dual differently. 

Substituting the above form of perturbed Elko in the scalar
perturbed Einstein equations lead to the following set of
equations:
{\small
\begin{subequations}
\label{eq:Pert}
\br
\label{eq:Pert00}
&& \Delta\Psi- 3 \Hm \Psi'-(\Hm'+2\Hm^2[1+\Fpt(\bphi)])\Psi  \\
& & = \frac{1}{2 \MPl} [\bphi'\dphi'+a^2 V_{,\varphi}\dphi] 
+ 3 \Fpt (\bphi)\Hm [\Psi'-\frac{\Hm}{\bphi}\dphi]  \nn \\
\label{eq:Pert0i}
&& \Psi'+\Hm[1+\Fpt(\bphi)]\Psi = \frac{1}{2 \MPl} \, \bphi'\dphi  \\
\label{eq:Pertii}
& & \Psi'' + 3\Hm\Psi' +( \Hm'+ 2\Hm^2[1+\Fpt(\bphi)])\Psi \\
&& = \frac{1}{2 \MPl} [\bphi'\dphi'- a^2\frac{V_{,\varphi}}{2}\dphi]- 
\Fpt(\bphi)\Hm[\Psi'-\frac{\Hm}{\bphi}\dphi] \nn \\
& & \dphi'' - \Delta \dphi - \bphi'\l[4 - 3 \l(1 -\varepsilon\r) 
\Fpteps - 3 \sqrt{\Fpteps}\r] \Psi' \nn \\
& & + \Hm \left[2 + 3 (1-\varepsilon) 
\Fpteps + 2 \Fpt \right] \dphi' 
+ a^2\left[V_{,\varphi}\Psi+\frac{1}{2}V_{,\varphi \varphi}\dphi\right]
\nn  \\
\label{eq:Pertconde}
& & - \frac{3}{4}\Hm^2 \left[1 
- \frac{8}{3} \Fpt \l[3 + \frac{\cal G}{{\cal H} \overline{\varphi}'} 
- \delta \r] 
+ 4 (1-\varepsilon) \sqrt{\Fpteps} \right]\dphi  \\
& & \qquad \qquad  - 2 {\cal H} \overline{\varphi}' 
\l[3 +  \frac{\cal G}{{\cal H} \overline{\varphi}'} - \delta \r] \Psi
+\frac{2}{\sqrt{3}}\frac{\bphi'}{\Hm} \Fpteps
\nabla \Psi'= 0 \nn 
\er
\end{subequations}
}
where $\Fpteps = \Fpt/({\Fpt + \varepsilon})$.

\indent The above equations are the first key results
of this letter, regarding which we would like to stress the following
points:
(i) To our knowledge, this is the first time cosmological perturbation
equations have been derived for any spinor field. For simplicity, we
have rewritten the above equations in terms of slow-roll parameter
($\varepsilon$) and have not assumed the condition that
$\varepsilon \ll 1$ i.e. these equations are exact.
(ii) These equations clearly show that the effective perturbations of
the Elko can be represented as a single scalar degree of freedom
($\delta \varphi$). Note that as in canonical scalar field inflation,
$(\Psi, \delta \varphi)$ are not independent and related by
(\ref{eq:Pert0i}).
(iii) Eq. (\ref{eq:const2}) indicates that the perturbed condensate
introduces a preferred symmetry.  However, it does not appear in the
linear perturbation equations (\ref{eq:Pert}). This is similar to the
unperturbed ansatz used by Boehmer \cite{Boehmer2007i,Boehmer2008}
which did not affect the background dynamics, but which, breaks the
isotropy in the perturbations.
(iv) As shown by Eq. (\ref{eq:Pertconde}), the sound speed of
perturbations for the Elko is identical to that of the canonical
scalar field.
(v) In the slow-roll limit, the non-adiabatic part of the
perturbations vanishes on super-Hubble scales i.e., the entropy
perturbation $\propto \nabla^2 \Psi$.
(vi) The terms $\varepsilon\sqrt{\Fpteps}, \Fpteps$ are of the same
order\footnote{For the choice $\epsilon_{_{\rm can}}\!\!\sim 0.1, \Fpt
\sim 0.01$, we get $\sqrt{\Fpteps} \epsilon\!\sim 0.03, \Fpteps \sim
0.1$.} as $\varepsilon_{_{\rm can}}$. These terms introduce non-trivial 
corrections to Mukhanov-Sasaki (MS) equation.

\par
\noindent {\it Power spectra:} The scalar perturbation equations 
(\ref{eq:Pert00}-\ref{eq:Pertconde}) are complicated and, as mentioned earlier, $(\Psi,
\delta \varphi)$ are not independent. In order to extract the physical
content and obtain the scalar power spectra transparently, it is
necessary to rewrite these equations in terms of the MS variable ($Q$)
which is linearly related to the curvature perturbation ${\cal R}$
\cite{Kodama-Sasa:1984}. Unlike the canonical scalar field, 
here it is not possible to obtain the MS variable using the 
gauge transformation properties of the metric and spinors. 
However, it is possible to obtain ${\cal R}$. Using the fact that 
$Q$ is related to ${\cal R}$, the MS variable for the Elko condensate, 
with an overall factor, is given by:
{\small
\beq
\label{eq:MSdef}
Q =  a \, \dphi + z \, \Psi \quad \mbox{where} \quad 
z = \left[1- \Fpteps \right] (a \bphi')/\Hm \, .
\eeq
}
Substituting $Q$ and $z$ in Eqs. (\ref{eq:Pert}), the MS equation is:
{\small
\beq
\label{eq:MSequation}
Q'' - \l[\nabla^2  + \frac{z''}{z} 
        - \ln[1-\Fpteps]'' 
        + \frac{7 \Hm' \Fpteps^{\frac{1}{2}} }{2}
        + \frac{\Hm \varepsilon' 
       \Fpteps^{\frac{1}{2}}}{\varepsilon}\r] Q \simeq 0 
\eeq
}

The above equation is the second key result of this letter. We 
would like to stress the following points:
(i) We have ignored the contributions of higher-order slow-roll
functions like $\varepsilon^2 \Fpteps^{1/2}, \Fpteps$. In the leading
order slow-roll approximation, their contribution to the scalar power
spectrum is tiny.
(ii) The Elko modification to the canonical MS equation is determined
by $\Fpt$.
(iii) Unlike the canonical scalar field, Eq. (\ref{eq:MSequation})
contains terms other than $z''/z$. This is due to the fact that we
have obtained $Q$ and $z$ by relating to ${\cal R}$, and not by using
the gauge transformation properties.
 
Invoking the slow-roll conditions ($\varepsilon, |\delta| \ll 1$) in the
scalar (\ref{eq:MSequation}) and tensor perturbation equations
(\ref{eq:TensorPertu}) --- and following the standard quantization
procedure by assuming Bunch-Davies vacuum --- the scalar and tensor
power spectra are given by:
{\small
\br
{\cal P}_S(k) & =& \l(\frac{H^2}{8 \MPl \pi^2}\r)
\left(\frac{\varepsilon + \Fpt}{\varepsilon^2}\right) 
\l[1- 2(c_0+1)\varepsilon_{_{\rm can}} \r.
\nn \\
\label{eq:SPS}
& & 
\l. + 2\varepsilon_{_{\rm can}} x 
+ \frac{1}{6} \left[13 \varepsilon_{_{\rm can}}  - 
              12 \delta_{_{\rm can}}\right](x-c_0) \r] \\
\label{eq:TPS}
{\cal P}_T(k) &=& \l(\frac{2 H^2}{\MPl \pi^2}\r) 
\l[1- 2(c_0+1) \varepsilon_{_{\rm can}} + 2 \varepsilon_{_{\rm can}} x\r]
\er
}
\hspace*{-7pt} 
where $c_0=\gamma_{_{\rm Euler}}+ \ln2- 2$ is a constant, $x=\ln(k^*/k)$
and $k^*$ is the pivot scale.

Eqs. (\ref{eq:SPS}, \ref{eq:TPS}) allow us to draw important
conclusions which are the third key result of this letter. Firstly,
the scalar and tensor power spectra of the spinflaton, in
slow-roll, are nearly scale-invariant. The scalar and tensor spectral
indices, expressed in terms of $\varepsilon_{_{\rm can}},
\delta_{_{\rm can}}$, are
{\small
\begin{subequations}
\begin{eqnarray}
\label{eq:nS}
n_{_S}&=& 1 - 4 \varepsilon_{_{\rm can}}
\l[1 +\frac{\sqrt{\Fpteps}}{4} - \Fpteps \r] 
+ 2\delta_{_{\rm can}}(1-\Fpteps) \\
\label{eq:nT}
n_{_T} &=& - 2 \varepsilon_{_{\rm can}}\l[1-(\Fpteps-\Fpt)\r] \, .
\end{eqnarray}
\end{subequations}
}
\hspace*{-7pt} Note that $\varepsilon_{_{\rm can}} \Fpteps^{1/2}$ and 
$\varepsilon_{_{\rm can}}$ have similar contributions. Secondly, in the 
leading order of $\epsilon_{_{\rm can}}$, the 
running of spectral indices are non-zero and given by 
{\small
\begin{subequations}
\br
\frac{d n_S}{d \ln{k}} &=&
-\frac{\varepsilon_{_{\rm can}}}{2}- 4 \varepsilon_{_{\rm can}}
\Fpteps^{1/2} + \frac{\varepsilon_{_{\rm can}}}{2}\frac{\Fpt}{1+\Fpt} \\
\frac{d n_T}{d \ln{k}}&=& 2 \, \varepsilon_{_{\rm can}} \, \Fpteps^{1/2}\, . 
\er
\end{subequations}
}
\hspace*{-7pt} It is interesting to note that the running of scalar spectral 
index ($-0.09 < d n_S/d (\ln k) < 0.019$) is consistent with the
WMAP-5 year results \cite{Komatsu2008}. For instance, using the WMAP
value of $\varepsilon_{_{\rm can}} = 0.038$ \cite{Komatsu2008} and
assuming that $\Fpt$ is tiny, we get $d n_S/d (\ln k) \sim
-0.019$. This is the one of the main predictions of spinor driven inflation.
Other observational implications of the running will be discussed in
Ref. \cite{DamienShanki:2008b}. 

Thirdly, the tensor-to-scalar ratio $r$ is no longer equal to $16
\varepsilon_{_{\rm can}}$ and is given by:
\beq
\label{eq:r}
r \simeq 16 \, \varepsilon_{_{\rm can}} \, \l[1- 2 \Fpteps\r] \, .
\eeq
Physically, this suggests that the gravitational wave contribution
during slow-roll spinor driven inflation is smaller than for canonical inflation. 
Lastly, as for the canonical scalar field, the scalar and
tensor perturbations during spinor driven inflation originate from the scalar
condensate and they are not independent. Hence, consistency
relations link them together. The one which is observationally useful
is the relation between $n_{_T}$ and $r$. Using Eqs. (\ref{eq:nT},
\ref{eq:r}), we get:
{\small
\beq
n_{_T}=\frac{r}{8} (1+\Fpteps) \l[ 1  
               + \varepsilon_{_{\rm can}}\l[\frac{11}{6} c + \Fpteps-\Fpt \r] 
               - 2\delta_{_{\rm can}} \, c \r] \, .
\eeq
}
\indent To conclude, for the first time, we have studied the linear
cosmological the metric perturbations and the perturbations of the
spinorial part of the field. With in the frame work of the Elko
forming a scalar condensate, we have shown that spinflaton can lead to
at least $60-$e foldings of inflation and that the primordial spectra,
in the slow-roll limit, are nearly scale invariant. We have shown that
the spinor driven inflation predicts a running of scalar spectral
index consistent with WMAP-5 year data. We have also shown that the
consistency relation has a non-trivial relation between $n_{_T}, r,
\Fpt, \varepsilon_{_{\rm can}}$ and $\delta_{_{\rm can}}$.

\vspace*{5pt}

\noindent {\it Acknowledgment}: The authors wish to thank Christian G. Boehmer,
Roy Maartens and L. Sriramkumar for several discussions. SS is
supported by the Marie Curie Incoming International Grant
IIF-2006-039205.

\end{document}